\documentclass{aa}  

\usepackage{graphicx}
\usepackage{txfonts}

\usepackage{natbib}
\usepackage{natbib,twoopt}
\usepackage[breaklinks=true]{hyperref} 
\bibpunct{(}{)}{;}{a}{}{,} 

\begin{document}

   \title{Relativistic reflection in the average X-ray spectrum of AGN in the V\'eron-Cetty \& V\'eron catalogue}


   \author{S. Falocco
          \inst{1,2}
          \and
          F. J. Carrera\inst{2}
\and
          X. Barcons \inst{2}
\and
          G. Miniutti \inst{3}
\and
          A. Corral \inst{4}
          }

\institute{University Federico II, Via Cinthia, Building 6, 80126 Naples, Italy \\
  \email{falocco@fisica.unina.it}
   \and Instituto de F\'isica de Cantabria 
              (CSIC-UC), Avenida de los Castros, 39005 Santander, Spain 
 \and Centro de Astrobiolog\'ia (CSIC-INTA), Dep. de Astrof\'isica; ESAC, PO Box 78, E-28691 Villanueva de la Ca\~nada, Madrid, Spain
\and  Institute for Astronomy, Astrophysics, Space Applications and Remote
Sensing (IAASARS), National Observatory of Athens (NOA), Palaia Penteli,
15236, Athens, Greece
       }

   \date{Received 7 October 2013; accepted 16 June 2014}

 
  \abstract
   {The X-ray spectra of active galactic nuclei (AGN) unveil properties of matter around the super massive black hole (SMBH). }
   {We investigate the X-ray spectra of AGN focusing on Compton reflection and fluorescence. These are two of the most important processes of interaction between primary radiation and circum-nuclear material that is located far away from the SMBH, as indicated by the unresolved spectral emission lines (most notably the Fe line) in the X-ray spectra of AGN. Contributions from the inner accretion disk, affected by relativistic effects as expected, have also been detected in several cases. }
   {We studied the average X-ray spectrum of a sample of 263 X-ray unabsorbed AGN that yield 419023 counts in the 2-12 keV rest-frame band distributed among 388 \textsl{XMM-Newton} spectra. 
}
   {
We fitted the average spectrum using a (basically) unabsorbed power law (representing the primary radiation). From a second model that represents the interaction (through Compton reflection and fluorescence) of this primary radiation with matter located far away from the central engine (e.g. the putative torus), we found that it was very significantly detected. Finally, we added a contribution from interaction with neutral material in the accretion disk close to the central SMBH, which is therefore smeared by relativistic effects, which improved the fit at 6 sigma.
 The reflection factors are 0.65 for the accretion disk and 0.25 for the torus.
Replacing the neutral disk-reflection with low-ionisation disk reflection, also relativistically smeared, fits the data equally well, suggesting that we do not find evidence for a significant ionisation of the accretion disk.
}
   {
We detect distant neutral reflection associated with a narrow Fe line in the average spectrum of unabsorbed
 AGN with $\langle z \rangle$=0.8. Adding the disk-reflection component associated with a relativistic Fe line improves the data description at 6 sigma confidence level, suggesting that both reflection
 components are present.
The disk-reflection component accounts for about 70 \% of the
 total reflected flux. }


   \maketitle
%

\section{Introduction}
Super massive black holes (SMBH) reside at the centre of active galactic nuclei (AGN), gravity being the ultimate power of their luminosities. 
 
X-ray primary radiation can interact with circum-nuclear matter through Compton scattering, producing a significant spectral feature at E$>8$ keV, and fluorescent lines, the most intense of which is the iron K $\alpha$ line at 6.4 keV, as explained in \cite{reynolds2003} and references therein.
It is well-known that two components can be detected in the iron line: a narrow core that is always detected in the X-ray spectra of AGN,
and a broad component significantly detected in a fraction of sources, although its ubiquitous presence cannot be discarded \citep{bhayani}.
The narrow core is emitted from the torus and/or the broad line region, hereafter called BLR \citep{nandra2006,shu2010,shu2011}, while the broad component is emitted from the accretion disk and appears to be broadened and deformed mainly by relativistic effects, including gravitational redshift \citep{tanaka1995,fabian2000,nandra2007,delacalle2010}.


Iron lines from accretion disks open the interesting possibility of exploring how matter accretes onto the SMBH: the broad iron line in principle helps constraining the inner and outer radii of the disk, the inclination angle between the disk and the line of sight and, last but not least, it also provides lower limits on the BH spin \citep{reynolds2003,reynolds2008,brenneman2009,risaliti2013}.
The first results in this field nursed the expectation that it might be possible to use iron lines for this type of measurement: \texttt{ASCA} spectra unveiled broad and redshifted iron lines in AGN \citep{tanaka1995}. Later observations of
AGN confirmed the broad lines in some AGN but not
in others, which left a complex situation, even for well-exposed AGN observations \citep{musho95,chiang,yaqoob2001}. 
 The non-universal profiles of iron lines detected in AGN are probably due to different contributions from the accretion disk, the BLR, the molecular torus, or even from material farther away \citep{iwasawa2003,guainazzi2012}. 

Averaging techniques 
  were introduced \citep{nandra2007} with the aim of detecting the iron line cores and the 
relativistic components in large AGN samples with high significance to increase the count statistics. These methods were expected to be powerful tools for separating the two contributions and unveil the physical meaning behind them. By studying the integrated \textsl{XMM-Newton} spectrum of the Lockmann Hole field, \cite{streblyanska} detected an intense relativistic line. Subsequent works based on \textsl{XMM-Newton} observations of a variety of target fields \citep{corral2008,chaudhary2010,chaudhary2012} and on the deep \textsl{Chandra} fields \citep{falocco2012,brusa2005,iwasawa2011} have detected the narrow line core with high significance, but did not find much evidence for a relativistic line, for example the upper limit of the EW of the broad iron line was 300 eV for \cite{corral2008}. 
 \cite{falocco2013} stacked of the \textsl{XMM-Newton} 
observation in the Chandra deep Field South, which helped constraining the properties of the iron line, although it has
been difficult to separate the narrow-line core and the relativistic component.
Summarising, previous results based on averaging techniques could neither confirm nor exclude a broad Fe line with a typical, averaged EW of $\sim$100 eV as observed in local AGN with high-quality X-ray spectra \citep{delacalle2010}.

 It is the purpose of this paper to separate the two components in average iron K $\alpha$ lines by simultaneously modelling each component of the iron line together with its associated Compton reflection. In Sect. \ref{sample} we explore the properties of the sample, drawn from fairly good quality X-ray spectra from \textsl{XMM-Newton} of AGN and quasi-stellar objects (QSO) from the V\'eron-Cetty \& V\'eron catalogue. In Sect. 3 we describe the method. Sect. 4 describes the results which we discuss in the Sect. 5. Finally, the conclusions are presented in Sect. 6, and a consistency check of the results is given in the appendix.

Throughout this paper, we adopt the
cosmological parameters, $H_{\ 0}$=70 km s$ ^{-1}$ Mpc$^{-1}$, $\Omega_{matter}$=0.3 and 
$\Omega_{\Lambda}$=0.7. All the counts
refer to the net number of counts between 2 and 12 keV rest-frame; the signal-to-noise ratio (hereafter, S/R) is also calculated in the same band, as are all the luminosities between 2 and 10 keV in the rest-frame band, unless explicitly stated otherwise. 
The 2 - 12 keV rest-frame band is chosen to characterise the spectral quality (S/R and counts) reached by the individual spectra in the same energy window as investigated in the average spectrum (as explained in Sect. 4.2, the average spectrum is fitted between 2 and 12 keV in the rest-frame band).
The 2-10 keV rest-frame band is used for the luminosity estimates to allow a direct comparison between the luminosity-redshift coverage of our new sample and those involved in previous similar works \citep{streblyanska,corral2008,chaudhary2010,chaudhary2012,falocco2012,falocco2013}.

We used \texttt{xspec v. 12.5}
\citep{arnaud} for the X-ray spectral analysis.
All the estimated statistical errors in this paper correspond to the 90\% confidence level for one interesting parameter, unless specified differently.
\section{X-ray data}\label{sample}
\subsection{Parent surveys}
We explore the X-ray spectral properties of the AGN sample composed using the 2nd \textsl{XMM-Newton} Serendipitous Source Catalogue, hereafter 2XMM \citep{watson} and the V\'eron-Cetty \& V\'eron catalogue, hereafter VCV \citep{vcv,vcv2010}.
The V\'eron-Cetty \& V\'eron catalogue provides a very large and reliable compilation of AGN and QSOs, including vital information like the redshift; this catalogue has provided us with the largest parent sample for the type of averaging X-ray spectral analysis we conducted.

The 2XMMi-DR3 \footnote{http://xmmssc-www.star.le.ac.uk/Catalogue/2XMMi-DR3cat-v1.0.fits.gz}
 catalogue is a compilation of X-ray selected sources, including active galaxies, clusters of galaxies, interacting compact binaries, and active stellar coronae, with a total of 262902 unique X-ray sources. The median flux in the total energy band (0.2 - 12 keV) of the catalogue detections is ~ 2.5 10$^{-14}$ erg cm$^{-2}$ s$^{-1}$; in the soft energy band (0.2 - 2 keV) it is ~ 5.6 10$^{-15}$ erg cm$^{-2}$ s$^{-1}$, and in the hard band (2 - 12 keV) it is ~ 1.4 10$^{-14}$ erg cm$^{-2}$ s$^{-1}$. About 20\% of the sources have total fluxes below 10$^{-14}$ erg cm$^{-2}$ s$^{-1}$.

We pre-selected the sources by focusing on sources with a galactic latitude higher than 20$^{\circ}$ with good spectral quality, selecting the observations with good EPIC-pn data. In this step we used several flags of the catalogue: \texttt{SUM-FLAG=0} (the value range between 0 and 4 indicates the quality of the detection: the value is set to zero if none of the quality flags is set to true), \texttt{PN-FILTER}$\neq$UNDEF (at least one of the filters has been used in the pn detector), \texttt{PN-FLAG=FALSE} (no flag is allowed to have been set in the pn detector, to avoid possible problems for a detection), \texttt{PN-ONTIME}$\neq$0 (the total good exposure time of the CCD where the detection is positioned must be different than zero). 

The VCV catalogue contains 168940 unique sources, with redshift information available for 99\% of the sample. \cite{vcv2010} have flagged as bad redshifts estimated from low-dispersion slitless spectra because they are less accurate or even incorrect (because the emission lines may have been misidentified). All the other sources have been flagged as having secure redshift estimates. For this paper, we used only sources with redshifts marked as "secure"
  in \cite{vcv2010} and in other papers on their redshift estimates, for example \cite{savage84} and \cite{abazajian2003}, except for a single source that is marked as insecure in \cite{vcv2010}, but not on the original reference \citep{savage84}. 
We include under the generic AGN denomination all objects classified in the VCV catalogue as either AGN or QSOs. However, we do not include sources classified as low-ionization nuclear emission-line regions (LINER) or BL Lacertae (BL Lacs) to ensure that our sample does not contain sources whose X-ray emission is dominated by synchrotron radiation and not by radiative processes on the accretion disk and corona.
 The catalogue provides a classification into type 1 and type 2 objects according to the presence/absence of broad optical lines. 

The matching criterion adopted to cross-correlate the 2XMM with the VCV catalogue is that the angular distance between the X-ray source (coordinates in the 2XMM) and the optical source (coordinates in the VCV catalogue) must be less than 5 arcsec. 
We extracted all the spectra from the EPIC-pn and the EPIC-MOS cameras with more than 50 counts between 0.2 and 12. keV obs-frame (for the EPIC-MOS camera, we extracted the spectra with a total number of counts of MOS1 + MOS2 higher than 50). 

The count threshold in the broad 0.2-12~keV band was chosen to allow a minimum spectral quality. This cut ensures that we included those individual spectra that allow at least a very basic fit of the continuum and prevents us from introducing noisy sources in the analysis.

\subsection{Spectral extraction}

 The spectra were automatically extracted 
 for each detection and EPIC camera following the procedure of \cite{mateos2005}, which we briefly summarise below. For this task, we used the SAS software (version 9.0.0) and the XMM-CCF, version 258 (07 October 2009). 

XMM-Newton data of observations processed before 2010 were
affected by calibration problems that result in an energy shift of Fe spectral
features (e.g. see the release note of CCF 271). 
To assess whether our results were severely affected by this instrumental error, we treated MOS and pn data separately (we applied the same method, including the simulations, to each individual spectrum) and made consistency checks of our results (see the appendix).
 For the spectral extraction, we used circular extraction regions centred at the source positions.
The size of the
source extraction regions was selected to optimise the S/R via
the \texttt{SAS} task \texttt{eregionanalyse}. The extraction radius of the spectra varies from source to source, but typically ranges
between 14 and 20 arcsec.

 Background spectra were obtained in circular regions after
masking out all the detected sources.
The code automatically choise the background region after generating 16 background regions. It then flagged the background regions as bad according to the following criteria: regions falling into two or more chips,  without a sufficient (at least 90\%) number of good pixels ('good' pixels are those without nearby sources), or falling in a different chip than the source. If more than one background region was selected with these criteria, the code finally chose the one closest to the source. \\
Calibration matrices (\texttt{arf} and \texttt{rmf})
were obtained for each spectrum with the \texttt{SAS} tasks \texttt{arfgen} and \texttt{rmfgen}.

To maximise the S/R, we combined the MOS1 and MOS2 spectra and the corresponding response matrices when they were
observed using the same filters and with the same observing modes, and when the difference in off-axis angles was less than 1 arcmin.
Merged source and background spectra were obtained by summing the
individual spectra. 'Back-scale' values (size of the regions used to
extract the spectra) and calibration matrices for the combined spectra
were obtained by weighting the input data with the exposure times. For sources with more than one observation, we added all the MOS and
all the PN spectra and response matrices, again only when observed using the same filters and observing modes, and when the difference in off-axis angles was less than 1 arcmin. EPIC-MOS and EPIC-pn data were not merged because
of their very different responses.  After extraction and merging
(of the two MOS cameras and of multiple observations of the same
sources, when applicable), we proceeded to the spectral
analysis of the individual spectra.
\subsection{Sample selection}
 We fitted each individual spectrum in the 1 - 12 keV rest-frame band using a simple absorbed power-law (\texttt{pha*zpha*pow}). Since we needed to estimate the photon index and the column density of the absorber, we included the band between 1 and 2 keV to be sensitive to moderate absorption.
The energy bins below 1 keV had to be ignored to avoid any soft-excess contribution \citep{porquet2004}.
In the fit, we fixed the column density of the \texttt{pha} component to its Galactic value in the direction of the source. Instead, the column density of the \texttt{zpha} component was left free to vary; this represents the intrinsic absorption that we wish to characterise. Summarising, we estimated (for each individual spectrum) the value of the column density of the absorber at the source redshift, the power-law photon index, the power-law normalisation, and finally the luminosities between 2 and 10 keV (corrected for Galactic and intrinsic absorption). The goodness of the fits to the individual spectra was verified with a $\chi^2$ test. The median value of $\chi^2_\nu$ = 53.6/48 indicates that this simple parametrisation seems to be a sufficient description of the individual spectra.

A number of cuts and down-selections were applied to the result of the cross-correlation to guarantee that our sample is composed of high-quality X-ray spectra of clean (unabsorbed) AGN and that the results are not biased in any conceivable way. 

To begin with, we kept only sources with fewer than 10000 spectral counts in the 2-12 keV rest-frame band: this prevents the very few spectra with high counts from dominating our analysis. We excluded sources with luminosities below $10^{40}$erg/s (to keep only bonafide AGN). 
 
Then, we further down-selected 672 spectra from EPIC-pn and EPIC-MOS cameras (401 unique AGN) with an S/R higher than 15. 

Finally, we only focused on bonafide 'unabsorbed' AGN, selecting objects for which the upper end of the 90\% error bar on the intrinsic column density $N_{\rm H}$ is $<10^{21.5}$~cm$^2$. 
Since reflection-dominated sources sometimes mimic unabsorbed AGN with very flat slopes, we have also excluded sources for which the lower end of the  $\Gamma$ is $<1$ at 90 \% confidence level. 

\begin{figure}
\centering
\includegraphics[width=8.3cm,angle=0]{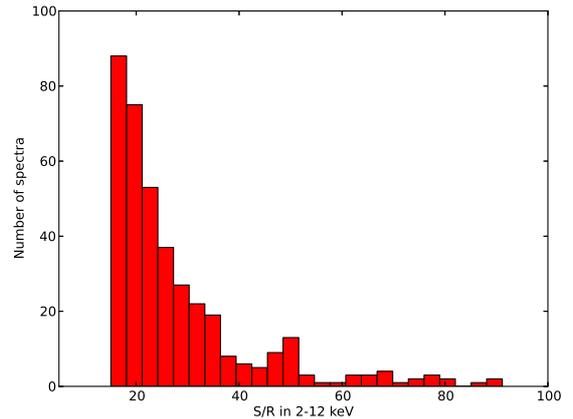}
\caption{Histogram of the S/R of the spectra used in this work.}\label{Fig:1}
  \end{figure}

 Our selected sample contains a total of 388 spectra (see Table \ref{tproperties}) from the EPIC cameras -MOS and/or pn-. Hereafter, our analysis is focused only on this sample, which contains 263 unique AGN with a total of 419023 counts in the 2-12 keV rest-frame band. Fig. \ref{Fig:1} displays the S/R distribution of the spectra in our selected sample.

 \begin{figure}
\centering
\includegraphics[width=8.3cm,angle=0]{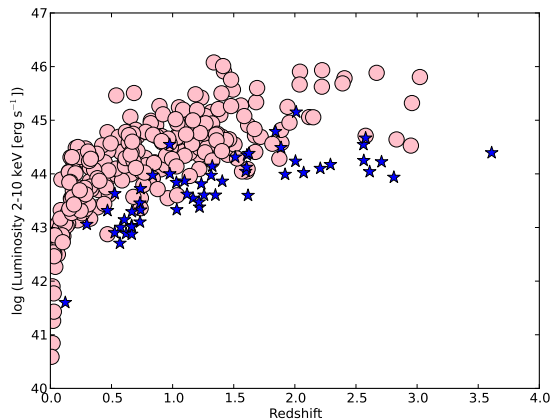}
\caption{Distribution of the selected sample in the luminosity and redshift plane: the circles represent the sources presented in this paper and the stars the AGN in the XMM CDFS studied in \cite{falocco2013}.}\label{Fig:2}
  \end{figure}

Fig. \ref{Fig:2} shows the distribution of the selected sample in the luminosity (of the continuum, from the fits to the individual spectra, corrected for Galactic and intrinsic absorption) and redshift plane of the sources. The same plot shows, for comparison, the sources of the XMM CDFS that were studied in \cite{falocco2013}. We note that the new sample provides more sources at every redshift and that the new sources have higher luminosities at all redshifts.


\begin{table*}

\centering
  \caption{ Properties of the selected sample.}
\centering
\begin{tabular}{ll}
 Number of spectra  & 388 \\
 Number of sources & 263  \\
 Net counts (2-12 keV rest-frame)  & 419023  \\
 Net counts (5-8 keV rest-frame) & 75598  \\
Average redshift  & 0.80  \\
 $\langle \log(L) \rangle $ $^1$ & 44.24 erg s$^{-1}$ \\
 $ \langle N_{\rm H,22} \rangle $  $^2$    & 0.01 $\times 10^{22}$ cm$^{-2}$\\ 
$\langle S/R \rangle$ $^3$  &28.02  \\

\end{tabular}

\tablefoot{ $^1$ Logarithm of the average  rest-frame 2-10~keV luminosity in erg s$^{-1}$, corrected for Galactic and intrinsic absorption; $^2$ average intrinsic column density; $^3$ average S/R of the sample between 2 and 12 keV rest-frame.
\label{tproperties}}
\end{table*}

\section{Method}\label{method}
To compute the average spectrum, we applied a method similar to that of \cite{corral2008}, as developed in \cite{falocco2013}, to the spectra in physical units. The procedure consists of the following steps: correcting the individual non-grouped spectra for the instrument energy-dependent response (this unfolding process is conducted by applying the best fit models obtained in the fits to the individual binned spectra, see Sect. 2.3), correcting for Galactic absorption, shifting to rest-frame, normalising according to the flux of the continuum, and finally, computing the unweighted mean (calculating the flux errors from the dispersion of the fluxes around the average).
It is known that the unfolding process can affect the results because it can introduce distortions into the individual spectra and the average spectrum. We characterised these effects in detail in \cite{falocco2012}, finding that for energies above 2 keV, the main effect of this procedure is to intrinsically broaden the iron line. For this reason, we study and characterise this spectral dispersion with simulations of unresolved lines (see below). The unfolding process can also affect the average observed and simulated continuum, but these effects are minimised if the best-fit continuum of each individual spectrum is used to unfold the observed and simulated spectra, as explained in \cite{falocco2012}. For this reason,
we estimated the expected shape of the underlying continuum in our spectra using extensive simulations.  
To do this, we simulated each individual spectrum and its background 110 times. We made use of the 'fakeit' command in Xspec for each spectrum. This tool produces the simulated source spectrum and its corresponding simulated background. The input model is the best-fit
absorbed power-law model from the individual fits to the observed spectra
 (see Sect. 2.1). To each of the simulated
samples we applied the same procedure as was used for the observed sample, thus obtaining 110 simulated average spectra. After this, we represented our simulated continuum with the median of the average simulated spectra.

As discussed at the beginning of this section, the spectral dispersion of the X-ray detectors as well as of the process performed (correction for the spectral response and the averaging process itself) can broaden any unresolved spectral features, and this spectral dispersion is energy dependent. 
To properly model spectral lines and absorption edges in the average observed spectra (e.g. the iron line as 6.4 keV or the Fe edge at 7 keV), we need to estimate our effective spectral dispersion in the energy band we analyse \citep{falocco2012}. 
To do this, we ran high S/R simulations of unresolved ($\sigma=0$) Gaussian emission lines for each spectrum. These are centred at several rest-frame energies between 1 and 10 keV, in steps of 1 keV. In addition, we simulated also a rest-frame 6.4 keV unresolved line. 
 The input model was a simple power-law with $\Gamma=1.9$ (no intrinsic absorption) plus an unresolved line with EW=200 eV (one simulation for each source -at its own redshift- and for each centroid energy). We corrected the simulated spectra for detector response (modelling them with a power-law with photon index 1.9); then we applied the same process to the synthetic X-ray spectra, in the same way as to the observed spectra. 
From these simulations we recovered the input parameters of the continuum very well, reproducing a power-law continuum with $\Gamma \sim$1.9, while the Gaussian lines appear to be broadened by $\Sigma\sim$ 100 eV. 
A similar value of the spectral dispersion given by the stacking procedure and the X-ray instruments was found in the study of the \textsl{XMM-Newton} X-ray spectra in the CDFS sample \citep{falocco2013} and in the COSMOS sample \citep{iwasawa2011}. 
 We underline that the values of the $\Sigma$ obtained in this work include contributions from the spectral dispersion of the instruments: the EPIC camera has a resolution of $\Sigma \sim$ 55 eV at 6 keV \citep{haberl2002}, which means that the averaging process introduces a very significant dispersion in spectral features, from 55 to 100 eV.

\begin{figure}
\centering
\includegraphics[width=8.3cm,angle=360]{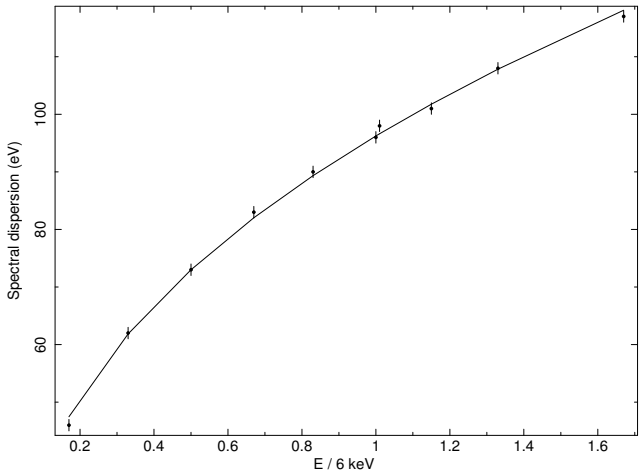}
\caption{Trend of the spectral resolution $\Sigma$ as a function of the spectral energy: $\Sigma =~96 \rm eV ~\left(\frac{E}{6~ \rm keV}\right)^{0.401}$ for the selected sample. }\label{Fig:3}
  \end{figure}

We studied the width of the average simulated lines as a function of their centroid energies, and we can see this trend in Fig. \ref{Fig:3}. A power-law fit to the points shown in Fig. \ref{Fig:3} yields $\Sigma =~96 ~\ \rm eV ~\left(\frac{E}{6~ \rm keV}\right)^{0.4}$.


 \begin{figure}
\centering
\includegraphics[width=6cm,angle=270]{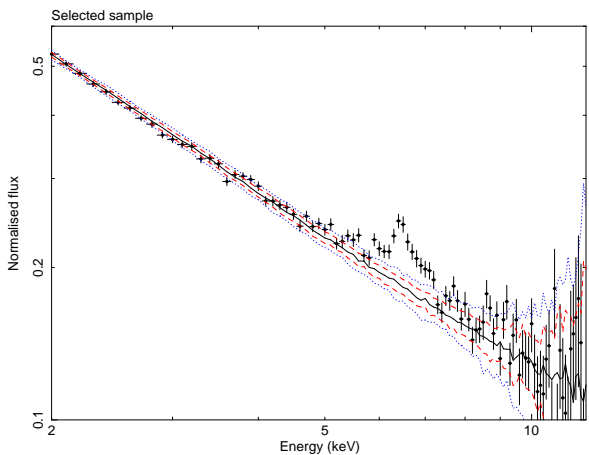}
\caption{Average spectrum. The data points represent the data, the continuous line the simulated continuum, the red dashed lines and the green dotted lines represent the one- and two-sigma confidence lines of the simulations. }\label{Fig:4}
  \end{figure}

 This effect is accounted for in the spectral fits by convolving the models used in the fits of the average spectra with a Gaussian smoothing (see Sect. \ref{modeldependent}).

\section{Results}\label{results}
\subsection{Model-independent study of the iron line}\label{modelindependent}
The average observed spectrum is shown in Fig. \ref{Fig:4}, were the continuous line represents the simulated continuum.  one- and two-sigma confidence lines of the simulated continuum are computed from the 68\% and 95\% percentiles around the median. 
 
A broad excess around the 6.4 keV rest-frame energy can be seen in Fig. \ref{Fig:4} and is quantified through our model-independent method in this section and the spectral fitting described in the next section. 
An excess also characterises the observed continuum between 8 and 10 keV. Its absence from the average simulated spectrum means that it can be interpreted as a property of the source spectra and not as an instrumental artefact due to the background (as specified in the previous Section, both the background and the source spectra have been simulated). For this reason, this continuum feature is better investigated with the spectral fitting in the next section.

We computed the EW of the iron line using the average simulated
spectra \citep{falocco2013} as follows: for each simulation we calculated the EW as
EW=$ \int_{E_1}^{E_2} dE \frac{T(E)-C(E)}{C(E)}\sim  \sum \Delta (E_i)\frac{T(E_{i})-C(E_{i})}{C(E_{i})}$,
where $\Delta(E_i)$ is the width of the spectral bin centred at $E_i$; $T(E_i)$ represents the average observed spectrum, $C(E_i)$ the average simulated continuum.
The values of the median of the EW for a number of energy windows are shown in Table \ref{teqw}. The confidence levels in that table are calculated from the 68\% percentiles around the corresponding medians.

We considered
several energy intervals, designed to describe a broad asymmetrical
iron line (5.5-7 keV), a symmetrically broadened (6.-6.8 keV) and
narrow (6.2-6.6 keV) iron line centred at 6.4 keV, finally, we
investigated an unresolved line with sigma approximately equal to the instrumental
resolution (estimated in Sect. \ref{method}), centred at 6.4 keV
and 6.9 keV, to represent emission from neutral and ionised iron,
respectively.

\begin{table}
 \caption{Iron line EW of the selected sample, in eV}
\scalebox{0.90}{
\begin{tabular}{rrrrrr}
 Band & \emph{Bulk} & \emph{Neutral} &\emph{Narrow} &\emph{Unresolved} &\emph{Ionised} \\
Band (keV) & 5.5-7. & 6.-6.8 & 6.2-6.6 & 6.3-6.5 
& 6.8-7.\\
\hline
EW (eV) \vline &  $131 \pm12 $
 & $95  \pm9$
&    $63 \pm7 $
&  $44 \pm6 $
& $26 \pm6$ 
\end{tabular}
}
 \tablefoot{EW expressed in eV, calculated from the simulations (see text) in different bands expressed in the first line, in keV.}\label{teqw}
\end{table}


From Table \ref{teqw}, we can see that higher EW characterise the iron feature for broader ranges, which suggests, at least qualitatively, the presence of a broad iron line. The significance of this feature is characterised in a model-dependent way in the next section. 

\subsection{Modelling the iron line}\label{modeldependent}
We fitted the average spectrum using models that take into account
direct radiation of the central engine and its reprocessing on
circum-nuclear material. Under the most simple assumption this
material is located outside the BLR, in the torus (or farther away),
and produces Compton reflection and fluorescence (mainly the iron
K$_{\alpha}$ line at 6.4 keV). We fitted the average rest-frame spectrum between 2 and 12 keV.
For the average redshift of our sample z$\sim$0.8 (see Table 1) the 10 keV rest-frame energy would correspond to the 5.5 keV observed frame energy, which are the corresponding energies for the redshifts that include 68\% of our sample (z$\sim$0.2 and 1.4) 4.2 and 8.3 keV (respectively), which is well within the sensitivity range of the instruments on board XMM-Newton.
 Including such high energies is not expected to strongly affect the average observed spectrum, and it is necessary to enable a good characterisation of the broad X-ray band continuum that also includes the Compton reflection. 

To account for the Compton reflection we used the \texttt{pexrav}
model in \texttt{xspec}. This model \citep{magdziarz1995}
is an exponentially cut-off power-law spectrum reflected from neutral
material. The model does not consider any fluorescent line, for which
we used a narrow Gaussian centred at 6.4 keV. We used the model
\texttt{gsmo*pha*(pexrav+pow+gauss)} (\emph{model 1} in Table
\ref{results}), which allowed us to obtain an EW estimate that can be directly
compared with previous works.  The \texttt{pha} component refers to the intrinsic absorption because the fit is performed in rest-frame and Galactic absorption has already been taken into account and corrected for.  
 In \emph{model 1}, the power-law represents
the primary emission, while \texttt{pexrav} and \texttt{gaussian}
account for its reprocessing far from the central SMBH (e.g. in the
torus). We used the following settings in the model: element abundances with respect to those in the interstellar medium of our galaxy $A$ $\equiv$
$A_{Fe}$ $\equiv1$, inclination angle $\theta$ $\equiv$30 and reflection factor $R_{non-rel}$ left free to vary (we first fitted the
spectrum excluding 5.5-7 keV, with \texttt{gsmo*pha*(pexrav+pow)}, then
added the narrow Gaussian at 6.4 keV using the model
\texttt{gauss}). Hereafter, we use the absolute value of the reflection component strength (in \texttt{xspec} models it has a negative value when only the reflected component is computed and then added to the already accounted-for continuum). 

\begin{figure}
\includegraphics[width=6cm,angle=270]{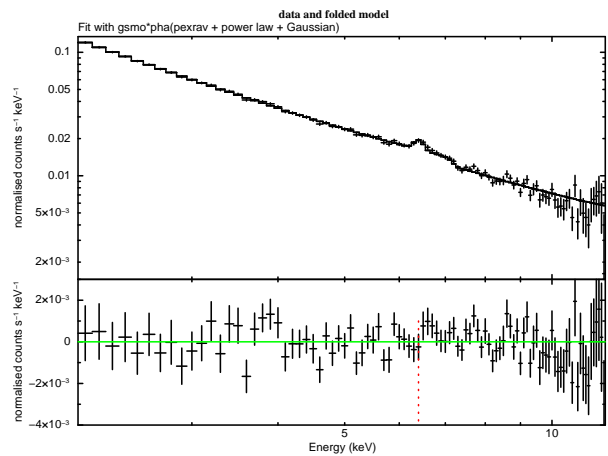}
\caption{Average spectrum of the selected sample fitted with \emph{model 1}. The bottom window displays the residuals between the data and the model. } \label{Fig:model1}
  \end{figure}

The first row of Table \ref{results} shows the fits with \emph{model 1}. We
were unable to constrain the values of the iron abundance
or to the inclination angle, because the error bars were too high.  The main
results obtained with \emph{model 1} are our constrains on the column
density, which confirms the unabsorbed nature of AGN included in this
analysis (as we can see from Table \ref{results}), on the reflection
factor $R_{non-rel}$= 4$^{+2}_{-1}$, and on
the iron line, with EW = 53$\pm7$ eV. The value of $R_{non-rel}$
found here is physically inconsistent with the low EW of the narrow line \citep{george1991}. This possibly indicates that the reflection model is trying to account for some additional spectral curvature (e.g. a broad Fe line).
 With \emph{model 1}, we were able to
constrain the Compton reflection and the fluorescent K$\alpha$ line: the error
bars at 90\% in Table \ref{results} indicate that we separated the contribution from the continuum and from the iron
line. This shows that our data require both components.
Moreover, from Fig. \ref{Fig:model1} we can see that the spectrum still presents some residual which indicates that other components are probably required as well.  

As mentioned before, we used \emph{model 1} with the
purpose of estimating the iron line EW and directly compare our results with previous works. When this was achieved, we used \texttt{pexmon},
which combines Compton reflection and fluorescence in a physically consistent manner. 

The model \texttt{pexmon},
developed by \cite{nandra2007}, uses a \texttt{pexrav} and fluorescent
lines: Fe K $_{\alpha}$, Fe K $_{\beta}$, Ni K $_{\alpha}$, and the Fe K
$_{\alpha}$ Compton shoulder. All these spectral components are
produced in a neutral material. This model allows us
to check whether the line strength is compatible with the continuum
reflection strength in the framework of the reflection model and with the largely unabsorbed nature of the sources
included in this sample.

\begin{figure}
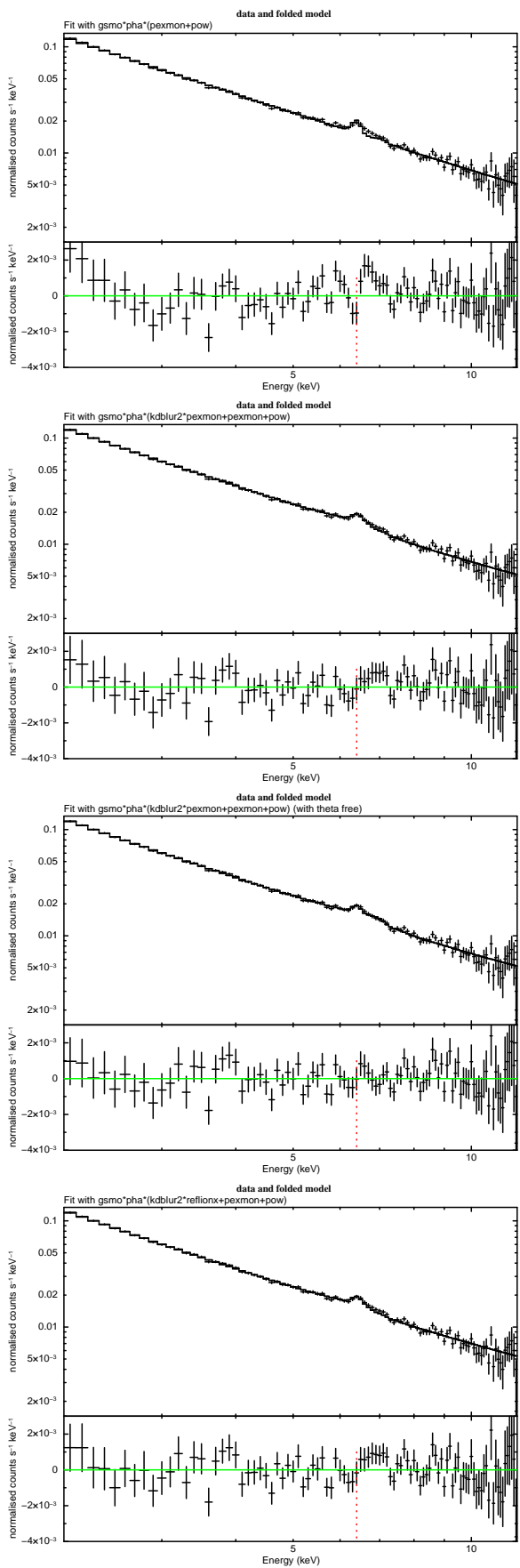

\includegraphics[width=5.8cm,angle=270]{figures/unabs_gam_ma1_gsmo_pexm_pow_res.ps}\\
\includegraphics[width=5.8cm,angle=270]{figures/unabs_gam_ma1_gsmo_pexm_kdblur_pexm_pow_res.ps}\\
\includegraphics[width=5.8cm,angle=270]{figures/unabs_gam_ma1_gsmo_pexm_kdblur_pexm_pow_tfree_res.ps}\\
\includegraphics[width=5.8cm,angle=270]{figures/unabs_gam_ma1_gsmo_pexm_kdblur_reflex_pow_res.ps}
\caption{Average spectrum fitted with \emph{model 2} (top panel), \emph{model 3} (in the second panel with fixed angle and in the third panel with free angle), and \emph{model 4} (bottom panel).  The bottom window of each panel displays the residuals between the data and the model.}\label{Fig:pexmon}
  \end{figure}

The model we used to fit our spectrum is in \texttt{xspec}:
\texttt{gsmo*pha*(pexmon+pow)} and is called also \emph{model
  2}. Similarly to the previous model, the power law accounts for the
direct component. This
model with one single \texttt{pexmon} component considers reprocessing in regions far
away from the central engine, most probably in the torus.  Using
\emph{model 2}, we found a reflection factor $R_{non-rel}=0.46$, compatible with the
picture of moderately absorbed spectra. The fit has
$\chi^2/dof=144.5/96$ and the residuals between data and model in
Fig. \ref{Fig:pexmon} (top panel) still display some excess that we need
to take care of. The following step is made to overcome this
disagreement with the model used so far, by investigating whether there is an accretion disk component.

The next model considers reprocessing by matter both far away from (e.g. in the
torus) and close to (e.g. in the accretion disk) the central
engine. The spectrum reprocessed in the accretion disk is smeared by
relativistic effects, which we can in principle observe in the
innermost regions of the accretion disk. The model we used is
\texttt{gsmo*pha*(kdblur2*pexmon+pexmon+pow)}, hereafter called
\emph{model 3}. The power law is again the primary emission, the reflection fraction of the
first \texttt{pexmon} is called hereafter $R_{disk}$, and the reflection fraction of the second one is called $R_{non-rel}$ (this is the reflection component already considered in \emph{model 2}). \texttt{kdblur2} emissivity (radial) profile has a power-law index =
0 and 3, and the transition radius between the two emissivity laws is 20
$R_g$. The accretion disk extends from an inner radius of $R_{in}$=1.235 $R_g$ to an outer radius $R_{out}$= 400
$R_g$. This setting has been chosen because it is the same as was used in
\cite{falocco2013}, which allows a direct comparison with the previous
results (see Sect. \ref{discussion}). 

We obtained continuum parameters consistent with moderate
absorption,
resulting in R$_{non-rel}=0.25\pm0.09$ and
R$_{disk}=0.65\pm0.10$, see Table \ref{results}.  This model
improves the fit with a $\chi^2/dof=111.8/95$, although some residuals are still left,
as we can see in the second panel of Fig. \ref{Fig:pexmon}.  We thus
tried to let the iron abundance vary, but we did not find any significant
improvement (the improvement of the fit was marginal, $\chi^2/dof=109/94$), and we were unable to constrain its value. Next, we left the inclination
angle $\theta$ of the accretion disk free to vary (the $\theta$ of \texttt{kdblur2} and of \texttt{pexmon} are tied together and $\theta$ of material far from the SMBH is fixed as in the previous fit), and we obtained a significant improvement for an inclination angle of $\theta=42^{+6}_{-5}\ ^{\circ}$, as we
can see from the $\chi^2/dof=100.5/94$ in Table \ref{results} and the
residuals between data and model in Fig. \ref{Fig:pexmon}, third panel.

Focusing again on the second panel of Fig. \ref{Fig:pexmon}, the
residuals of the fits to \emph{model 2} show an excess between
6.4 and 7 keV, that might suggest fluorescence from ionised iron. To
account for any ionised contribution, we substituted in the fit the
\texttt{pexmon} smeared by relativistic effects, with a model that represented the same processes in ionised material. In particular, we used
one private version of the \texttt{reflionx} model that is available as an additional table 
model from the \texttt{xspec} web site \citep{ross2005}. This version of \texttt{reflionx} model allows the ionisation parameter ($\xi$ is defined as $4\pi\frac{F}{n}$, where F is the total illuminating flux and n is the hydrogen number density) to be $>$ 1 $\rm erg$ $\rm cm$ $\rm s^{-1}$, while the public version of \texttt{reflionx} allows it to be $>$30 $\rm erg$ $\rm cm$ $\rm s^{-1}$.
The total model is thus
\texttt{gsmo*pha*(kdblur2*reflionx+pexmon+pow)}, hereafter called
\emph{model 4}.  The new parameter introduced with \texttt{reflionx} is
the ionisation parameter, as we can see from the Table \ref{results},
constrained to be lower than 58 $\rm erg$ $\rm cm$ $\rm s^{-1}$ at a 90\% confidence level. A visual comparison between the fourth (model 4) and the first panel (model 2) of Fig. 6, as well as the results in Table \ref{results}, shows that model 4 much improves the spectral fitting. However, comparing the fourth panel with the second one (model 3, using neutral disk reflection) in the same figure, there is no improvement, meaning that both neutral and ionised disk reflection are equally suitable for the data modelling. This strongly suggests that a disk reflection component (neutral or ionised) is strongly required by the data, which are not very sensitive to the disk ionisation, however.

\begin{table*}
\caption{Fits of the average spectrum with model 1: \texttt{gsmo*pha*(pexrav+pow+gauss)}; model 2: \texttt{gsmo*pha*(pexmon+pow)}; model 3: \texttt{gsmo*pha*(kdblur2*pexmon+pexmon+pow)}; model 4: \texttt{gsmo*pha*(kdblur2*reflionx+pexmon+pow)}; \label{results} }
\begin{tabular}{rrrrrrrrrrr}
\hline\hline

Ref.& $\rm N_{H}$  & $\Gamma$  & $R_{non-rel}$ & $R_{disk}$  &  A$\equiv\rm A_{Fe}$  & $\theta$ & EW  &  Norm$_{ion}$  &  $\xi$   &$\chi^2/\nu$   \\
&($10^{22}\rm cm^{-2}$)  &     & &  & (A$_{\odot}$)  & ($\rm deg$) & (eV) & ($\frac{\rm photons}{\rm keV~ cm^{2}~ s} $)    & (${\rm erg} $ ${\rm cm}$ $ {\rm s}^{-1}$)   & \\

  (1)& (2)  & (3)  &  (4) & (5)  & (6) & (7) & (8) & (9) & (10) & (11) \\

1 & 5$\times10^{-3}(<0.82)$ & 2.00$_{-0.04}^{+0.08}$ & 4$_{-1}^{+2}$ & -  & $\equiv$1 &  $\equiv$30 & 52$\pm$7 & - & - & 109.0/94 \\





\hline\hline



2 &  $\equiv0$ & $1.87^{+0.01}_{-0.01}$  & $0.46\pm0.07$ & - & $\equiv$1 & $\equiv$30 &- & - & -  &  144.5/96 \\ 
\hline \hline

3 &  $\equiv0$ & $1.92^{+0.01}_{-0.02}$  & $0.25\pm0.09$ & $0.65_{-0.10}^{+0.10}$ & $\equiv$1 & $\equiv$30 &- & - &-  &  111.8/95\\ 

3 &  $\equiv0$ & $1.93^{+0.02}_{-0.01}$  & $0.31\pm0.06$ & $0.72_{-0.09}^{+0.08}$ & $\equiv$1 & $ 42.4^{+6.0}_{-5.0}$* &- & - & -  &  100.5/94\\



4 &  $\equiv0$ & $1.92^{+0.03}_{-0.03}$  & $0.28\pm0.10$ & - & $\equiv$1 & $\equiv$30 & - &$3^{+100}_{-2}\times10^{-4}$  & 33$^{+25}_{-33}$  &  113.7/94  \\




\end{tabular}
 \tablefoot{Columns: (1) reference of the fit model (see text); (2) column density of the intrinsic absorber; (3) photon index; (4)  reflection factor of the non-relativistic component; (5) reflection factor of the disk component; (6) abundances; (7) inclination angle; (8) EW of the non-relativistic line component; (9) normalisation of the \texttt{reflionx} component; (10) ionisation parameter; (11) chi-squared and number of degrees of freedom. *: the inclination angle refers to the kdblur2 and pexmon convolved by kdblur, while the inclination of the other pexmon is fixed to 30 degrees.}
\end{table*}

 \begin{figure}
\centering
\includegraphics[width=6.5cm,angle=270]{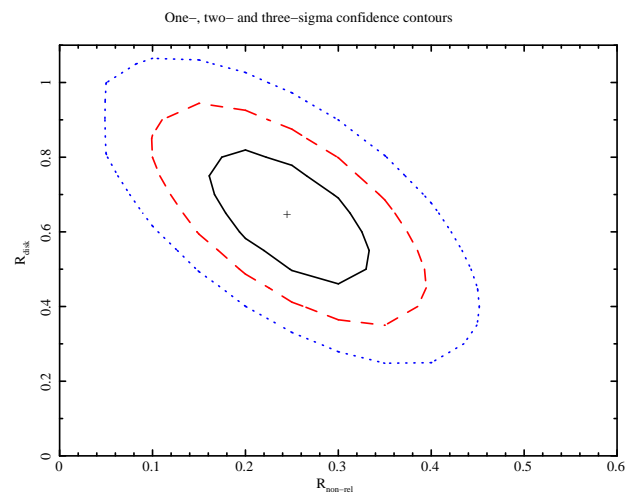}
\caption{
1- (solid line), 2- (dashed line), 3- (dotted line) sigma contours of the $R_{rel}$ versus $R_{non-rel}$ of \emph{model 3}}\label{Fig:contours}
  \end{figure}


\section{Discussion}\label{discussion}

We used several physical models to fit the iron line and the continuum detected in the average X-ray spectrum of the selected sample of VCV QSOs and AGN that are present in the 2XMM X-ray source catalogue.

The fit with \emph{model 1}, which makes use of Compton reflection and fluorescence as independent spectral components, was used with the aim to estimate the EW of the iron line, which is 52 eV, in broad agreement with previous works, for example \cite{nandra2006} and \cite{nandra2007}. The fit helps to understand that the spectral quality reached here allows us to separate the broad band continuum and the iron feature with high confidence. The 90 \% errors in Table \ref{results} indicate that the Compton reflection continuum and the fluorescent Fe line are significantly detected.
A reflection factor of $R=4^{+2}_{-1}$ is inconsistent for unabsorbed or moderately absorbed AGN, being more typical for reflection dominated AGN. Moreover, this value is inconsistent with the low EW of the narrow iron line \citep{george1991}.

For this reason, we next used a reflection model that accounts in a meaningful manner for Compton reflection and fluorescence, \emph{model 2} in Table \ref{results}. The value of R$_{non-rel}=0.46$ found with this fit is compatible with the moderately absorbed spectra that constitute the selected sample. The data, as shown from the $\chi^2/dof=144.5/96$ in Table \ref{results}, require a more complex model. For this reason we considered reprocessing (both fluorescent lines and Compton hump) both in the torus and in the accretion disk, \emph{model 3} in Table \ref{results}.

With \emph{model 3}, we constrained the reflection factors to be 0.65 for the disk component and 0.25 for the torus. We checked the confidence contours of the two parameters: Fig \ref{Fig:contours} shows the one-, two- and three-sigma confidence levels of the two $R$. These contours do not cross zero values, which means that both components are securely detected. The significance of the accretion disk spectral component is 6 sigma. 

Comparing this result with those of \cite{corral2008} and \cite{falocco2012}, we here introduce important novelties. \cite{corral2008} found a significant narrow iron line core and only an upper limit to the relativistic component (EW$<$300 keV) in the average spectrum of the XMS-XWAS sample. In the study of the deep \textsl{Chandra} Fields by \cite{falocco2012}, the iron feature is significant but unresolved. In that work, a broad component was marginally detected only in a sub-sample of AGN and only at an $\sim$2 $\sigma$ confidence level. In both \cite{corral2008} and \cite{falocco2012} the narrow core was highly significant and the relativistic line was not significantly detected. This work, in contrast, presents a six sigma detection of accretion disk features, which include a relativistic Fe line. It is interesting to understand the origin of the apparent difference between the results obtained in the present paper and in those of \cite{corral2008}, the two samples being based on \texttt{XMM-Newton} spectra and surveys with similar sensitivity. The reason that we detected a more significant relativistic spectral component than \cite{corral2008} is that our sample was defined by selecting spectra with S/R$>$15. With this, we effectively excluded many sources with low spectral quality and a limited number of X-ray counts. The individual spectra included in this sample have more than 247 counts in the 2-12 keV rest-frame band, while those included in \cite{corral2008} have at least 80 counts (individually) in the same band. Low-count (and S/R) spectra increase the dispersion on the continuum of the average spectrum (determined both through the simulations and through the spectral fitting), thus any faint broad line would be more difficult to detect. 

Compared with results obtained by \cite{falocco2013}, this work again delivers an unprecedented result. The average
spectrum in the XMM CDFS field, was fitted with a model similar to \emph{model} 3 of the present work, but we found that the disk component and the non-relativistic reprocessing component were strongly coupled. In contrast, our confidence contours in Fig. \ref{Fig:contours} show that there is no degeneracy between the two components and that they are both detected significantly. Moreover, the three-sigma confidence levels exclude the zero value for both reflection components.
 We also note that the disk X-ray features detected in the XMM CDFS at two sigma of
confidence level in \cite{falocco2013} are now significant at a six-sigma confidence level. 

Comparing our results with those of \cite{streblyanska}, we find that our analysis is different from the methodological point of view, because in this work the model used for the spectral fitting was convolved with a Gaussian smoothing that uses the spectral dispersion across the X-ray spectrum. Moreover, in \cite{streblyanska}, the data were grouped before computing the integrated spectrum. In that work, the final integrated spectrum was modelled with a simple power-law and a line model, while we employed a comprehensive model that considered Compton hump and fluorescence emerging from both accretion disk and torus. 
These are probably the reasons why the EW estimates shown in this paper in Table \ref{teqw} are lower than the values of EW$\sim$500 eV found in \cite{streblyanska}.
In this work, the EW of the relativistic component is 72$\pm$9 eV, while the EW of the non-relativistic component is 55$\pm$7 eV (corresponding to the fits with \emph{model 3} ignoring the spectral bins above the iron line region). This estimate is broadly consistent with the upper limit found in \cite{corral2008} and with EW estimates found for individual AGN with high S/R, for example \cite{nandra2007}. 

Our data do not appear to be sensitive to element abundances, but by leaving the inclination angle of the accretion disk free to vary we found $\theta=42_{-5}^{+6}$ $^{\circ}$ with a fit improvement of 3.4 sigma with respect to the fit with $\theta$ fixed to 30$^{\circ}$. 

We finally tested for ionised
material by substituting the \texttt{pexmon} component from the accretion
disk with a \texttt{reflionx} model (that is, \emph{model 4} in Table \ref{results}). This model considers the same mechanisms as were
represented by pexmon, but with the addition of the ionisation parameter which can reach 1 $\rm erg$ $\rm cm$ $\rm s^{-1}$. The
fit improvement with respect to the one single \texttt{pexmon} fit is significant, but not as good as that introduced by the neutral reflection model. This shows that the accretion disk spectral component, blurred by relativistic effects, can be equally fitted with mildly ionised material (in this case it is possible to obtain only an upper limit for the ionisation parameter: $<60$ $\rm erg$ $\rm cm$ $\rm s^{-1}$) and with neutral material.

Summarising, our fits show material far away from and close to the central SMBH without strong evidence that the latter is ionised.


\section{Conclusions}
We investigated the iron line in distant AGN by studying the average spectrum of a large sample. This was constructed through the cross-correlation of the 2XMM \citep{watson} sample and the VCV \citep{vcv,vcv2010}, with a mean redshift of $\langle z \rangle$=0.8. The sample studied here contains 263 unique AGN. We averaged the spectra using a procedure used previously for \textsl{XMM-Newton} and \textsl{Chandra} samples by \cite{corral2008} and \cite{falocco2012,falocco2013}. The most relevant fits to our average spectrum were made using the model \texttt{gsmo*pha*(kdblur2*pexm+pexm+pow)}, which uses a primary emission (power law) reprocessed in material located in two spatially separated regions, producing two reflection components. The first one, smeared by relativistic effects due to the strong gravitational field of the central SMBH (\texttt{kdblur2*pexmon}), is emitted in the accretion disk. The second one (\texttt{pexmon}), which is not convolved with the relativistic kernel, comes from regions far enough from the SMBH for the relativistic effects to be neglected. This is most plausibly the torus. Our main conclusions are that
\begin{itemize}
\item the average detected spectrum has a significant Compton-reflection component and an iron line at 6.4 keV,
\item Compton reflection and fluorescence come from both the torus and the accretion disk, and
\item the reprocessing material is neutral or consistent with neutral because the ionisation parameter of the disk-reflection component has a 
relatively low upper limit.
\end{itemize}
The average spectrum of AGN presented in this work for the first time enabled a comprehensive study of both iron line and Compton reflection from circum-nuclear material in AGN. Compton reflection and fluorescence were studied in detail in \cite{corral2008}, but the contribution from material close to the SMBH was not significantly detected.  In the present work, we clearly separated the contribution from material far from the central SMBH (e.g. the torus) and that from material close to the SMBH, the accretion disk. A relativistic line was detected in \cite{streblyanska}, but the continuum was modelled considering only the primary emission.
 This might have brought an EW estimate higher than those found in individual bright AGN in the literature. The values of the iron line EW found in this work with a model-independent method are instead consistent with those previously found in individual local AGN.  The relativistic line, whose detection is more difficult when the continuum is modelled considering Compton reflection, is now unveiled at a six-sigma confidence level in our average spectrum, and proves to be common amongst AGN in the Universe.

\begin{acknowledgements}
Financial support for this work was provided by the Spanish Ministry of Economy and Competitiveness, through rolling grants AYA2010-21490-C02-01, AYA2010-21490-C02-02 and AYA2012-31447.
SF for this research made use of the Super Computer Altamira which belongs to the Spanish Super Computing Network. 
 The research leading to these results has also received funding from the European Union Seventh Framework Programme (FP7/2007--2013) under grant n. 312789. 

   The results presented in this work results are based on observations obtained with XMM-Newton, an ESA science mission with instruments and contributions
directly funded by ESA member states and the USA (NASA).

 \end{acknowledgements}

\bibliographystyle{aa}
\bibliography{bibtex}

\begin{appendix} 
\section{average spectrum of MOS data}
The spectra were reduced with the SAS software (version 9.0.0) and the XMM-CCF, version 258, (07 October 2009). In the release note associated with the CCF version 271, it is explained that the XMM-Newton observations made before 2010 have had a calibration problem that shifts the iron line by a quantity that depends on the instrument and on the observation. 
Since this effect varies wildly in magnitude among all the pn observations, it is rather difficult to characterise for the pn spectra included in the survey. In the MOS observations, instead, it introduces a dispersion with r.m.s. $\sim$ 10 eV. For this reason, we have run a consistency check on our results by analysing the MOS data alone and by summing the 10 eV broadening in quadrature to the intrinsic broadening determined with the simulations of the iron line. The results (see Table \ref{results_mos}) are fully consistent with the MOS+pn analysis discussed in Sect. \ref{discussion}. The detection of the neutral disk component is significant at 3.5 sigma, given the chi-squared difference between models 3 and 2 of Table \ref{results_mos}. The addition of an ionised disk reflection is instead significant at 4.2 sigma, from the comparison between rows 4 and 2 in the same table.
 The larger uncertainties connected to the MOS data are due to the lower spectral quality and total counts involved (the MOS spectra are 205 and have in total 210931 and 38270 counts in the 2-12 and 5-8 keV rest-frame bands). The consistency of our results with those using both the MOS and pn cameras confirms their robustness against the calibration issues that affect the EPIC cameras.   

\begin{table*}
\caption{Fits of the average spectrum of MOS data with model 1: \texttt{gsmo*pha*(pexrav+pow+gauss)}, model 2: \texttt{gsmo*pha*(pexmon+pow)}, model 3: \texttt{gsmo*pha*(kdblur2*pexmon+pexmon+pow)}, and model 4: \texttt{gsmo*pha*(kdblur2*reflionx+pexmon+pow)}. \label{results_mos} }
\begin{tabular}{rrrrrrrrrrr}
\hline\hline
 
Ref.& $\rm N_{H}$  & $\Gamma$  & $R_{non-rel}$ & $R_{disk}$  &  A$\equiv\rm A_{Fe}$  & $\theta$ & EW  &  Norm$_{ion}$  &  $\xi$   &$\chi^2/\nu$   \\
&($10^{22}\rm cm^{-2}$)  &     & &  & (A$_{\odot}$)  & ($deg$) & (eV) & ($\frac{\rm photons}{\rm keV~ cm^{2}~ s} $)    & (${\rm erg} $ ${\rm cm}$ $ {\rm s}^{-1}$)   & \\

  (1)& (2)  & (3)  &  (4) & (5)  & (6) & (7) & (8) & (9) & (10) & (11) \\

1  & $0.060_{-0.059}^{+0.165}$  & $2.02 ^{+0.11} _{-0.08}$  & $4.3_{-1.7}^{+2.1}$  & - & $\equiv$1  & $\equiv$30 & 54$\pm$8 & - & - & 93.7/94 \\

2  & $\equiv$0  & 1.87 $\pm$ 0.01  & 0.47 $\pm$ 0.09  & -  & $\equiv$1 & $\equiv$30 & - & - &-  & 104.9/96 \\

3  & $\equiv$0  & 1.92 $\pm$ 0.03  & 0.30 $\pm$ 0.13  &  $0.51 ^{+0.26} _{-0.25}$ &$\equiv$1  & $\equiv$30 & - & - & - & 92.8/95 \\

4  & $\equiv$0  & 1.86 $\pm$0.02  & 0.30 $\pm$ 0.12  & -  &$\equiv$1  &$\equiv$30  & - &2.7$^{+15.0}_{-1.3}\times10^{-6}$  & $990_{-641}^{+901}$  & 83.76/94 \\

\end{tabular}
 \tablefoot{Columns: (1) reference of the fit model (see text); (2) column density of the intrinsic absorber; (3) photon index; (4)  reflection factor of the non-relativistic component; (5) reflection factor of the disk component; (6) abundances; (7) inclination angle; (8) EW of the non-relativistic line component; (9) normalisation of the \texttt{reflionx} component; (10) ionisation parameter; (11) chi-squared and number of degrees of freedom.}
\end{table*}

\end{appendix}

\end{document}